\documentclass[a4paper,12pt]{article}
\usepackage{amsmath,amssymb,amsthm}
\usepackage{geometry}
\usepackage{graphicx}
\usepackage[noend]{algpseudocode}
\usepackage{algorithm}
\usepackage{braket}
\usepackage{hyperref}
\usepackage{booktabs}
\usepackage{authblk}
\usepackage{tablefootnote}
\usepackage[backend=biber,style=numeric]{biblatex}

\title{A Quantum Genetic Algorithm Framework \\For The MaxCut Problem}
\author{Paulo A. Viana\thanks{CIn, UFPE. Email: pava@cin.ufpe.br} \and Fernando M de Paula Neto\thanks{CIn, UFPE. Email: fernando@cin.ufpe.br}}

\date{December, 2024}

\begin{document}

\maketitle

\label{chap:metodologia}

\begin{abstract}
The MaxCut problem is a fundamental problem in Combinatorial Optimization, with significant implications across diverse domains such as logistics, network design, and statistical physics. The algorithm represents innovative approaches that balance theoretical rigor with practical scalability. The proposed method introduces a Quantum Genetic Algorithm (QGA) using a Grover-based evolutionary framework and divide-and-conquer principles. By partitioning graphs into manageable subgraphs, optimizing each independently, and applying graph contraction to merge the solutions, the method exploits the inherent binary symmetry of MaxCut to ensure computational efficiency and robust approximation performance. Theoretical analysis establishes a foundation for the efficiency of the algorithm, while empirical evaluations provide quantitative evidence of its effectiveness. On complete graphs, the proposed method consistently achieves the true optimal MaxCut values, outperforming the Semidefinite Programming (SDP) approach, which provides up to 99.7\% of the optimal solution for larger graphs. On Erd\H{o}s-R\'{e}nyi random graphs, the QGA demonstrates competitive performance, achieving median solutions within 92-96\% of the SDP results. These results showcase the potential of the QGA framework to deliver competitive solutions, even under heuristic constraints, while demonstrating its promise for scalability as quantum hardware evolves.
\end{abstract}

\section{Introduction}

The MaxCut problem is a classical combinatorial optimization challenge where the goal is to partition the vertices of a graph into two disjoint subsets such that the sum of the weights of the edges between the subsets is maximized. Formally, given a graph \( G = (V, E) \) with edge weights \( w(e) \), the objective is to find a subset \( S \subseteq V \) that maximizes:

\[
\text{Cut}(S) = \sum_{(u, v) \in E, u \in S, v \notin S} w(u, v).
\]

MaxCut has practical applications in data clustering \cite{Poland_Zeugmann_2006}, statistical physics, and circuit layout design. For a comprehensive survey, check \cite{Poljak_Tuza_1995}. However, due to its NP-hard nature, exact solutions are computationally expensive, especially for large graphs. 
The best known algorithm for polynomial cost approximation is the method proposed by Goemans and Williamson using semidefinite programming (SDP) and randomized rounding, which achieves an approximation of 0.878 of the best answer \cite{goemans1995improved}.
Heuristic methods using artificial intelligence have been used in search of approximate solutions, such as genetic algorithms \cite{kim2019comparison} and neural networks \cite{wang2006improved, jiang2021more}.


Quantum computing has algorithms that surpass the theoretical limits of known classical algorithms, by promoting effects such as entanglement, superposition, and phase cancelation \cite{martin2022quantum,jozsa2003role}. The Deutsch-Jozsa quantum algorithm \cite{deutsch1992rapid} verifies whether a given function is balanced or constant, in constant time; Grover's algorithm \cite{grover1996fast} performs a search in a disordered structure and has a quadratic gain in relation to the linear cost of the best known classical algorithm; as well as Shor's algorithm \cite{shor1994algorithms}, which factors numbers with polynomial cost, presenting an exponential gain in relation to the best known classical algorithm.
The use of artificial intelligence in quantum algorithms has allowed the construction of quantum models that learn from data\cite{biamonte2017quantum}, as well as the execution of intelligent models in a quantum architecture that promotes efficiency in their execution \cite{lahoz2016quantum}.

This paper proposes a Quantum Genetic Algorithm (QGA) using an evolutionary framework based on Grover’s algorithm and divide-and-conquer principles. By partitioning graphs into manageable subgraphs, optimizing each one independently, and applying graph contraction to merge the solutions, the method exploits the inherent $\mathbf{Z_2}$ symmetry of MaxCut to ensure computational efficiency and robust approximation performance. The results are compared with the classical SDP model. 

The paper is organized as follows. In Section~\ref{sec:TF}, the basic concepts to understand the proposed model are presented, which are principles of quantum computing, operation of Grover's quantum Algorithm, Genetic Algorithm and Quantum Genetic Algorithm, and details of graphs. In Section~\ref{sec:MaxCutProblem}, the MaxCut problem is formally presented and some discussions about its known complexity limits are given. In Section~\ref{sec:QGAFramework}, the proposed framework based on a quantum genetic algorithm, Grover's quantum algorithm, and the divide-and-conquer technique is presented. Also in Section~\ref{sec:QGAFramework}, the complexity analysis of the proposed model is presented. In Section~\ref{sec:experiments}, experiments are presented that validate the proposed model. In Section~\ref{sec:analysis}, an analysis and discussion of the results found are presented. In Section~\ref{sec:conclusion}, the conclusions of the paper are presented.


\section{Theoretical Foundations}
\label{sec:TF}

\subsection{Quantum computing}

A quantum bit, or qubit, is a unit vector in a two-dimensional complex vector space. Although the basis states for a qubit can be chosen from any orthogonal basis of $\mathbb{C}^2$, the most commonly used basis is the $\textit{canonical}$ (or $\textit{computational}$) basis. This basis is defined by the pair of vectors $\ket{0} = [1,0]^T$ and $\ket{1} = [0,1]^T$. A qubit $\ket{\psi}$ can be expressed as shown in Equation~\ref{eq:qubit}, where $\alpha$ and $\beta$ are complex numbers satisfying the normalization condition $|\alpha|^2 + |\beta|^2 = 1$.

\begin{equation} \label{eq:qubit} \ket{\psi} = \alpha\ket{0} + \beta\ket{1} \end{equation}

In quantum computing, it is fundamentally impossible to create an exact copy of an unknown quantum state, as stated in the no-cloning theorem \cite{nielsen:00}. Composite quantum systems are described using the tensor product, such that $\ket{ij} = \ket{i} \otimes \ket{j}$. Typically, quantum computing involves Hilbert spaces of dimension $2^k$, where $k$ represents the number of qubits. A general state of $k$ qubits can be written as:

\begin{equation} \label{eq:k-qubit} \ket{\phi} = \sum_{i=0}^{2^k-1} \alpha_i \ket{\mathbf{i}} \end{equation}

It is standard practice to use binary notation for the basis states $\ket{\mathbf{i}}$. For example, a $4$-qubit basis state $\ket{\mathbf{5}}$ can be represented in binary form as $\ket{0101}$ or equivalently as $\ket{0} \otimes \ket{1} \otimes \ket{0} \otimes \ket{1}$. Consequently, it is meaningful to refer to specific qubits in a binary representation. For instance, the last qubit of the state $\ket{\mathbf{5}}$ is $\ket{1}$.

Transformations in a quantum system are achieved through operators, which modify the amplitude values of qubits. A quantum operator $\mathbf{U}$ acting on $n$ qubits is represented by a $2^n \times 2^n$ complex unitary matrix. A unitary matrix $\mathbf{U}$ satisfies the condition $\mathbf{U}^\dagger \mathbf{U} = \mathbf{U} \mathbf{U}^\dagger = \mathbf{I}$, where $\mathbf{U}^\dagger$ is the Hermitian conjugate (or conjugate transpose) of $\mathbf{U}$, and $\mathbf{I}$ denotes the identity matrix.

Some frequently used single-qubit operators include the \textit{NOT} operator $\mathbf{X}$ and the \textit{Hadamard} operator $\mathbf{H}$. These are defined as:

\begin{equation} \label{eq:quantumop1} \mathbf{X} = \begin{bmatrix} 0 & 1 \ 1 & 0 \end{bmatrix}, \quad \mathbf{H} = \frac{1}{\sqrt{2}} \begin{bmatrix} 1 & 1 \ 1 & -1 \end{bmatrix}. \end{equation}

Finally, a quantum circuit consists of a sequence of quantum operators applied to one or more qubits. Each operator represents a discrete step in the transformation of the quantum state.

\begin{equation}
\label{eq:quantumop1}
\begin{array}{ll}
\textbf{I}= \left[
\begin{array}{ll}
 1 & 0\\
 0 & 1\\
\end{array}
\right]
\begin{array}{l}
\textbf{I}\ket{0}= \ket{0} \\
\textbf{I}\ket{1}=\ket{1} 
\end{array}
\end{array}
\begin{array}{ll}
\textbf{X}= \left[
\begin{array}{ll}
 0 & 1\\
 1 & 0\\
\end{array}
\right]
\begin{array}{l}
\textbf{X}\ket{0}= \ket{1} \\
\textbf{X}\ket{1}=\ket{0} 
\end{array}
\end{array}
\end{equation}

\begin{equation}
\label{eq:quantumop2}
\begin{array}{ll}
\textbf{H}=\frac{1}{\sqrt{2}} \left[
\begin{array}{cc}
 1 & 1\\
 1 & -1\\
\end{array}
\right]
&
\begin{array}{l}
\textbf{H}\ket{0}= 1/\sqrt{2}(\ket{0}+\ket{1}) \\
\textbf{H}\ket{1}=1/\sqrt{2}(\ket{0}-\ket{1}) 
\end{array}
\end{array}
\end{equation}

The identity operator $\mathbf{I}$ acts trivially on the input state, leaving it unchanged. The NOT operator $\mathbf{X}$ functions analogously to the classical NOT gate, flipping the computational basis states. The Hadamard operator $\mathbf{H}$, on the other hand, generates a superposition of states by applying an equal-weighted transformation. The $\mathbf{CNOT}$ (controlled-NOT) operator, which takes two input qubits and produces two output qubits, flips the second qubit (the target qubit) if and only if the first qubit (the control qubit) is in the state $\ket{1}$. This behavior is illustrated schematically in Figure~\ref{fig:cnot}.

The $\mathbf{CNOT}$ operator is a two-qubit gate characterized by a control qubit and a target qubit. Its operation depends on the value of the control qubit: when the control qubit is in the state $\ket{1}$, the $\mathbf{X}$ operator is applied to the target qubit, flipping its state. If the control qubit is in the state $\ket{0}$, no operation is performed on the target qubit. The matrix representation of $\mathbf{CNOT}$ in the computational basis is shown in Equation~\ref{eq:CNOTworking}.

\[
\mathbf{CNOT} = \begin{bmatrix}
1 & 0 & 0 & 0 \\
0 & 1 & 0 & 0 \\
0 & 0 & 0 & 1 \\
0 & 0 & 1 & 0 \\
\end{bmatrix}
\]

\[\mathbf{CNOT}\ket{00}= \ket{00}\] 
\[\mathbf{CNOT}\ket{01}= \ket{01}\]
\[\mathbf{CNOT}\ket{10}= \ket{11}\]
\[\mathbf{CNOT}\ket{11}= \ket{10}\]

\subsection{Quantum Grover's Algorithm}

Grover's algorithm is a quantum algorithm designed to solve the problem of searching for a marked element in an unstructured database or solving a black-box function inversion problem. It provides a quadratic speedup over classical counterparts, reducing the search complexity from \(O(N)\) to \(O(\sqrt{N})\), where \(N\) is the size of the search space.

Let \(f: \{0, 1\}^n \to \{0, 1\}\) be a Boolean function such that:
\[
f(x) = 
\begin{cases} 
1, & \text{if } x = x^\ast, \\ 
0, & \text{otherwise}.
\end{cases}
\]

Here, \(x^\ast\) represents the single "marked" element. The goal is to find \(x^\ast\) with as few evaluations of \(f\) as possible. This problem can be easily converted to checking if there is a given in a disordered data structure \cite{giri2017review}.

Grover's algorithm uses quantum parallelism and amplitude amplification to locate \(x^\ast\). It involves the following steps: \textbf{State Initialization}, \textbf{Oracle Query}, \textbf{Amplitude Amplification} and \textbf{Iterative Search}. Below are the detailed each one step of Grover's Algorithm.

\textbf{State Initialization}

The quantum system is initialized to an equal superposition of all the basis states.

\begin{enumerate}
    \item Prepare the \(n\)-qubit system in the initial state:
    \[
    \ket{\psi_0} = \ket{0}^{\otimes n}.
    \]

    \item Apply the Hadamard transform \(\mathbf{H}^{\otimes n}\) to generate the uniform superposition:
    \[
    \ket{\psi_1} = \frac{1}{\sqrt{N}} \sum_{x=0}^{N-1} \ket{x}, \quad \text{where } N = 2^n.
    \]
\end{enumerate}

\textbf{Oracle Query}

The oracle \(\mathbf{O}\) is a quantum operator that flips the sign of the amplitude of the marked state \(\ket{x^\ast}\). Mathematically, it is defined as:
\[
\mathbf{O}\ket{x} = 
\begin{cases} 
-\ket{x}, & \text{if } x = x^\ast, \\ 
\ket{x}, & \text{otherwise}.
\end{cases}
\]

After applying the oracle, the quantum state becomes:
\[
\ket{\psi_2} = \frac{1}{\sqrt{N}} \left( \sum_{x \neq x^\ast} \ket{x} - \ket{x^\ast} \right).
\]

\textbf{Amplitude Amplification (Grover Diffusion Operator)} \\
Grover's diffuser operator, often called the inversion about the mean, is a key component of Grover's search algorithm. It amplifies the amplitudes of marked states (those satisfying the Oracle's condition) while reducing the amplitudes of unmarked states, effectively focusing the search on desired solutions. This section details the definition, implementation, and significance of Grover's diffuser operator.

The diffuser operator, denoted as $D$, performs the transformation:
\[
D = 2|s\rangle\langle s| - I,
\]
where:
\begin{itemize}
    \item $|s\rangle = \frac{1}{\sqrt{N}} \sum_{x=0}^{N-1} |x\rangle$: The equal superposition state.
    \item $I$: The identity operator.
    \item $N$: The total number of states in the search space.
\end{itemize}

This operation reflects the quantum state about the average amplitude of all states, enhancing the probability of measuring marked states. \\

The diffuser operator is implemented using the following steps:
\begin{enumerate}
    \item \textbf{Initialization to Superposition:}
    Apply a Hadamard gate to each qubit to create the equal superposition state $|s\rangle$ if not already prepared.
    
    \item \textbf{Phase Inversion:}
    
     The diffuser reflects the amplitudes of all states about their average, indirectly amplifying the marked states (those identified by the Oracle) through iterative applications. This is represented by the operator $-I$.
    
    \item \textbf{Reflection About Mean:}
    Reflect all states about the mean amplitude. This is achieved using:
    \[
    D = H^{\otimes n} (2|0\rangle\langle 0| - I) H^{\otimes n},
    \]
    where $H^{\otimes n}$ is the Hadamard gate applied to $n$ qubits.
\end{enumerate}

Generally, the diffuser can be broken down into the following steps:
\begin{enumerate}
    \item Compute the mean amplitude of all states.
    \item Invert the amplitude of each state about this mean.
\end{enumerate}
The next step increases the amplitude of the marked state using the Grover diffusion operator \(\mathbf{D}\), which reflects the amplitudes about their average value. The operator is defined as:
\[
\mathbf{D} = 2\ket{\psi}\bra{\psi} - \mathbf{I}.
\]

After applying the diffusion operator, the new state is given by:
\[
\ket{\psi_3} = \mathbf{D} (\mathbf{O} \ket{\psi_1}).
\]

\textbf{Iterative Search}

The combination of the oracle \(\mathbf{O}\) and the Grover diffusion operator \(\mathbf{D}\) is applied repeatedly. This combined operation is often called the \textbf{Grover operator} \(\mathbf{G}\), defined as:
\[
\mathbf{G} = \mathbf{D}\mathbf{O}.
\]

The state evolves iteratively as:
\[
\ket{\psi_{k+1}} = \mathbf{G}\ket{\psi_k}, \quad k = 0, 1, \dots, r-1.
\]

The number of iterations \(r\) required to maximize the probability of measuring the marked state is approximately:
\[
r = \left\lfloor \frac{\pi}{4} \sqrt{N} \right\rfloor.
\]

\textbf{Measurement}

After \(r\) iterations, the quantum state is dominated by the marked state \(\ket{x^\ast}\). A measurement in the computational basis yields \(x^\ast\) with high probability.

\subsection{Genetic Algorithm}

A Genetic Algorithm (GA) is a biologically inspired optimization technique that mimics the process of natural selection. It is a heuristic search method used for solving optimization and search problems by evolving a population of candidate solutions through iterations. In a GA, solutions are encoded as binary values called "chromosomes," and their quality is assessed using a fitness function. The algorithm consists of three main steps:

\begin{itemize}
    \item \textbf{Selection}: High-fitness individuals are selected to contribute to the next generation.
    \item \textbf{Crossover}: Selected individuals exchange parts of their genetic information to produce offspring.
    \item \textbf{Mutation}: Small, random changes are introduced to maintain diversity in the population and explore the search space.
\end{itemize}





\subsection{Quantum Genetic Algorithm}

Quantum Genetic Algorithm (QGA) tries to introduce the principles of genetic algorithms for quantum circuits to take advantage of quantum superposition in particular and quantum computing in general to obtain better results and performance. In a QGA:

\begin{itemize}
    \item The population of candidate solutions is represented as quantum states, enabling simultaneous evaluation of multiple solutions.
    \item Quantum gates are used to perform operations analogous to crossover and mutation in classical GAs, but these are implemented as reversible and unitary transformations.
    \item The fitness evaluation can utilize quantum circuits to identify and amplify high-quality solutions.
\end{itemize}

QGAs has in its aim the idea to enhance the exploration and exploitation capabilities of classical GAs by using quantum computing, leading to potentially faster convergence and better scalability for large search spaces.

\section{The MaxCut Problem}

\label{sec:MaxCutProblem}
\subsection{Theoretical Overview}
The MaxCut problem is a foundational optimization problem in graph theory with wide-ranging theoretical and practical implications. Formally, given an undirected graph $G = (V, E)$ with edge weights $w_{ij} \in \mathbb{R}^{+}$ for $(i, j) \in E$, the objective is to partition the vertex set $V$ into two subsets $(S, T)$ such that the sum of the weights of edges between $S$ and $T$ is maximized. This is mathematically expressed as:
\[
\text{Maximize } \sum_{(i,j) \in E} w_{ij}(1 - z_i z_j)/2,
\]
where $z_i \in \{-1, 1\}$ represents the partition assignment of vertex $i$.

The Max Cut problem is NP-hard and was one of the original NP-complete problems identified by Karp. It finds applications in various domains, including statistical physics, network clustering, and circuit design.

\subsubsection*{Lower Bound}
A natural lower bound for the Max Cut problem can be derived by considering a random partition of the vertices into two subsets. For any graph $G$, this approach ensures that each edge has an equal probability of being cut or not cut. Mathematically, the expected weight of a random cut is:
\[
\mathbb{E}[\text{Weight of Random Cut}] = \frac{1}{2} \sum_{(i,j) \in E} w_{ij}.
\]
This provides a baseline for evaluating approximation algorithms. Since a valid cut must have a weight at least equal to this expectation, the Max Cut problem satisfies the inequality:
\[
\text{Max Cut}(G) \geq \frac{1}{2} \sum_{(i,j) \in E} w_{ij}.
\]
This lower bound highlights that even the simplest random partition provides a meaningful starting point, achieving an approximation ratio of $1/2$ relative to the total edge weight.

\subsubsection*{Analytical Solution for Complete Graphs}
For a complete graph $K_n$, every vertex is connected to every other vertex. If all edges have uniform weight $w$, the Max Cut value can be determined analytically as \cite{edwards1973}:
\[
\text{Max Cut}(K_n) = \lfloor \frac{n^2 w}{4} \rfloor,
\]
where the partition divides the vertices as evenly as possible into two subsets. This solution arises because each edge contributes to the cut if and only if it spans the two subsets.

\subsubsection*{Binary Symmetry}
Since the goal is to maximize the sum of the edges between the subsets, exchanging the labels of the subsets does not alter the value of the solution. This implies that for any valid solution where a vertex is assigned to one subset, an equivalent solution exists by flipping all assignmented vertices. 

\subsection{Erd\H{o}s-R\'{e}nyi Random Graphs}
Common benchmarks used for testing approximation algorithms and exact-solution algorithms for the MaxCut include Complete Graphs, \textit{i.e.} graphs that each vertex is connected to all other vertices. And Erd\H{o}s-R\'{e}nyi Random Graphs.\\
The Erdős-Rényi random graph is a fundamental model in graph theory that generates random graphs through a probabilistic process. 
\textbf{\( G(n, p) \)}: In this model, a graph with \( n \) vertices is constructed by adding each possible edge between any two vertices independently with probability \( p \). This means the presence of each edge is determined randomly, leading to graphs with varying numbers of edges, though the expected number of edges is \( \binom{n}{2}p \).

The Erdős-Rényi model exhibits specific statistical properties. In the \( G(n, p) \) model, the degree of each vertex (the number of edges connected to it) follows a binomial distribution, which approximates a Poisson distribution for large \( n \) and small \( p \). The model also displays sharp transitions or threshold phenomena, where certain properties of the graph, such as connectivity or the emergence of a giant connected component, appear suddenly as \( p \) changes.

The clustering coefficient, which measures how likely neighbours of a vertex are to be connected, is equal to \( p \) in this model. For sufficiently large \( n \) and moderate \( p \), the diameter of the graph (the longest shortest path between two vertices) tends to grow logarithmically with \( n \), indicating that the graph remains relatively small in terms of overall distance between vertices.

The Erdős-Rényi model is a well-known method in the literature for understanding random graph behavior, and generating random graphs.
\section{Proposed QGA Framework}
\label{sec:QGAFramework}

The QGA usual framework has inherent flaws that are hard to deal with directly, however the work done by \cite{udrescu2006} showcases a novel approach into the implementation of a QGA, this framework is dubbed as the RQGA.This approach deals with the fitness values in parallel with its corresponding individual of the population, meaning that both the individual and its fitness value are in superposition simutaneously, making the need to make measurements to compute fitness values unnecessary. Moreover the frameworks gives a base to develop any new potential QGA based on a specific problem, one which was implemented by one of the authors \cite{ardelean2022}. Here the algorithm is distinctively implemented for the MaxCut algorithm, with the addition of a divide-and-conquer heuristic in form of graph contraction, inspired by \cite{zhou2023}). Firstly, the QGA for the MaxCut will be explained and afterwards the graph contraction heuristic.

\subsection{Standard QGA Framework}

The standard QGA operates as forcing the classical genetic algorithm components into a quantum framework. Its steps typically include:
\begin{enumerate}
    \item \textbf{Population Initialization:} Representing a population of individuals as quantum states. The population is encoded as a superposition of all possible solutions:
    \[
    |\psi_\text{pop}\rangle = \frac{1}{\sqrt{2^N}} \sum_{i=0}^{2^N - 1} |u_i\rangle,
    \]
    where $N$ is the number of qubits used to encode an individual.

    \item \textbf{Quantum Operators:} Applying quantum equivalents of classical genetic operators:
    \begin{itemize}
        \item \textit{Quantum Mutation:} Perturbing quantum states to explore the solution space.
        \item \textit{Quantum Crossover:} Combining features from multiple quantum states to create new solutions.
    \end{itemize}

    \item \textbf{Fitness Evaluation:} Measuring the fitness of each individual in the population by collapsing the superposition.

    \item \textbf{Selection:} Using quantum principles, such as amplitude amplification, to prioritize fitter individuals.

    \item \textbf{Iteration:} Repeating the process for a fixed number of generations or until convergence.
\end{enumerate}

\subsection{Enhanced Approach}

The enhanced framework deviates from the standard QGA by rethinking the reliance on traditional genetic operators. It introduces significant modifications to streamline the process:

\subsubsection{Overview}
\begin{figure}[h]
    \includegraphics[width=1\textwidth]{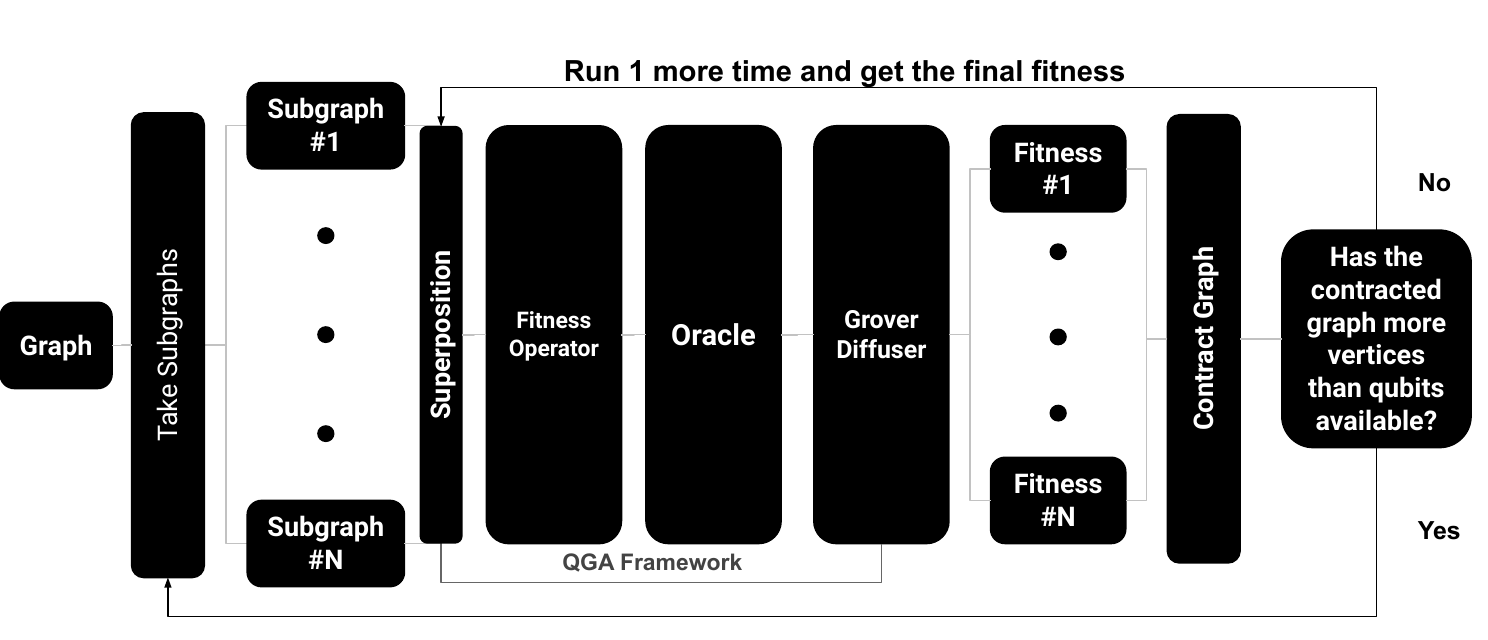}
    \caption{Diagram of the proposed framework.}
    \label{fig:diagram}
\end{figure}

The algorithm presented for the MaxCut begins by recursively partitioning the graph using the METIS library until each subgraph reaches a number of vertices less than or equal to a set limit determined by the number of qubits available. For each subgraph, the individual register for each vertex and the fitness register, which is limited by the number of edges in the graph, are created, and tThen, all of them are set in superposition, using quantum parallelism to represent the solution space of all possible results simultaneously. With this approach it is possible to explore all possible individual solutions and their corresponding fitness values simultaneously by marking them, eliminating the need for iterative genetic operations as in classical algorithms. A unitary operation representing the fitness function for the MaxCut is applied to compute the fitness of each individual, encoding these values into the fitness register. Subsequently, an oracle is applied to mark valid individuals whose values were set by the fitness operator, and Grover’s diffusion operator is used to amplify the amplitudes of these marked solutions. The system is then measured, collapsing the quantum state to one of the high-fitness individuals. This cycle of fitness computation, marking, amplitude amplification, and measurement is repeated until the threshold, theoretically limited by Grover's search and the nature of the MaxCut problem, ceases to improve, signaling convergence. Finally, the algorithm outputs the chromosome corresponding to the highest observed fitness value as the solution. The diagram of the process can be view on figure \ref{fig:diagram}.

\subsubsection{Superposition and Fitness Register}
Instead of relying on mutation and crossover, the enhanced approach encodes the entire population into a single quantum superposition, combining individual and fitness registers:
\[
|\psi\rangle = \frac{1}{\sqrt{2^N}} \sum_{i=0}^{2^N - 1} |u_i\rangle \otimes |0\rangle.
\]
An unitary operator $U_f$ based on a boolean function $f$ is applied to compute the fitness values:\newline
\[
U_f: |u\rangle \otimes |0\rangle \to |u\rangle \otimes |f(u)\rangle.
\]
This ensures the fitness values are calculated without requiring explicit genetic operators, reducing computational complexity.

\subsubsection{Optimization with Grover's Algorithm}
The enhanced framework reduces the fitness evaluation to a quantum maximum-finding problem using Grover's algorithm. By defining a specific oracle and employing amplitude amplification, it identifies the individual with the highest fitness value in $O(\sqrt{N})$ iterations, significantly outperforming classical exhaustive search methods.

\subsection{Key Differences}
Here we show the main differences of the proposed model compared with the standard framework found in the literature. While traditional QGAs rely heavily on adaptations of classical genetic operators to quantum computers, the model proposed introduces a more quantum-coherent approach. With the integration of the fitness values directly besides their respective individual using quantum parallelism and using Grover’s algorithm to amplify the best solutions, the proposed model overcomes several limitations of conventional QGAs for the MaxCut, like the dependency of computing angles and repetitive measurements. 

\begin{itemize}

    \item \textbf{Population Representation:} In standard QGAs, the population is represented solely as a superposition of individual states. Each state encodes a potential solution, but there is no direct association with fitness values within the quantum state. Conversely, the enhanced approach integrates the fitness information directly into the quantum state by combining individual and fitness registers. This integration allows simultaneous processing of both solution candidates and their fitness evaluations.\\

    \item \textbf{Genetic Operators:} The standard algorithms heavily rely on quantum analogues of classical genetic operators, such as mutation and crossover, to explore the solution space. These operations require iterative application to generate diverse solutions. The enhanced approach eliminates the need for these operators, since the solution is already established by the superposition, it utilizes Grover's algorithm to efficiently amplify the best solution's amplitude without requiring constantly reading the population to adjust circuit parameters. \\

    \item \textbf{Fitness Evaluation:} Fitness is typically evaluated by collapsing the quantum state and measuring each candidate solution, a process that requires repeated measurements to identify optimal solutions. The enhanced framework replaces this approach with a unitary operator that computes fitness values directly within the quantum superposition, avoiding the need for repeated state collapses. \\

    \item \textbf{Excluding Undesired Solutions:} There's no explicit mechanisms for distinguishing individuals within the population. This can lead to inefficiencies if undesired solutions are prioritized. The enhanced approach bi-partitions the fitness space into subspaces according to theoretical Lower Bound of the MaxCut value, ensuring that bad solution do not interfere with the optimization process. \\

    \item \textbf{Optimization Methodology:} Standard QGAs rely on iterative selection and modification processes, which can be computationally intensive. The enhanced approach employs a single-step optimization methodology using Diffusion operator of the Grover's algorithm to identify the best solution with quadratic speed-up compared to classical exhaustive search. \\

    \item \textbf{Computational Complexity:} The computational cost of standard QGAs depends on the number of iterations and the size of the population, often scaling poorly for large problems. In contrast, the enhanced framework achieves a complexity of $O(\sqrt{N})$, inheriting the quantum speed-up of the Grover's algorithm. \\
\end{itemize}
\begin{algorithm}
\caption{Enhanced Quantum Genetic Algorithm}
\begin{algorithmic}[1]
\State Initialize quantum registers: $|u\rangle$ (individuals) and $|f\rangle$ (fitness values)
\State Prepare a uniform superposition of all possible individuals:
\[ |\psi\rangle = \frac{1}{\sqrt{2^N}} \sum_{i=0}^{2^N - 1} |u_i\rangle \otimes |0\rangle \]
\State Apply the fitness function $U_f$ to compute fitness values:
\[ U_f: |u\rangle \otimes |0\rangle \to |u\rangle \otimes |f(u)\rangle \]
\State Define the oracle $O$ to mark states with the highest fitness values:
\[ O: |u\rangle \otimes |f\rangle \to (-1)^{g(f)} |u\rangle \otimes |f\rangle \]
\State Apply Grover's iterations:
\begin{enumerate}
    \item Apply the oracle $O$ to mark the highest fitness states.
    \item Perform the diffusion operator to amplify the marked states.
\end{enumerate}
\State Since Grover's algorithms requires $O(\sqrt{{N}}))$ queries to the oracle, where $N$ is the size of the search space. For a $M$-qubit fitness register, we have a search space of size $2^M$, and by knowing that for every graph we have 2 exact solutions for the problem, we would know that $ N = \frac{2^{M}}{2} = 2^{M-1}$. Giving us $\approx O(\sqrt{{2^{M-1}}}))$ necessary queries.
\State Measure the quantum state to obtain the individual with the highest fitness value.
\end{algorithmic}
\end{algorithm}

\subsection{Circuit Initialization}

The MaxCut problem requires the encoding of a \( |V| \)-vertices Graph cut solutions into a quantum register, where each binary combination represents a partition. 
The chromosome is represented as an \( (n \times |V|) \)-qubit quantum register, where \( |V| \) is the number of vertices and $n$ is the number of qubits used to represent the solution value. This number is determined \textit{a priori} by knowing that the maximum value of the cut is the number of Edges of the graph. Giving us that, for any graph with $|E|$ edges, $ n = \lceil \log_2{|E|} \rceil $.\\
Similarly, due to the theoretical knowledge of the MaxCut problem we can ignore all those solutions which have a cut less than $\frac{1}{2}$ of the total sum of the edges. So all those solutions are explicitly encoded in a subset which will be ignored, ensuring clear separation between the desired potential cut values and unneeded solutions.\\

\subsection{Fitness Subcircuit}
The fitness subcircuit computes the fitness values of the solutions and maps each individual solution to its corresponding fitness value encoded in a fitness register. \\
The fitness function \( f: \{G, K\} \to \mathbb{N}\) is defined as follows:
\begin{itemize}
    \item Returns \( 0 \), if the individual is undesired, \textit{i.e.}, it has its $MaxCut \leq \frac{1}{2}|E|.$
    \item Returns \( x \in \mathbb{N} \) , where \( x \) represents the number of edges with adjacent vertices belonging to different partitions.
\end{itemize}

The operator $U_{\text{fit}}$ is defined by the function $f$, performing the transformation:
\[
U_{\text{fit}}: |u\rangle \otimes |0\rangle \to |u\rangle \otimes |f(u)\rangle,
\]
where:
\begin{itemize}
    \item $|u\rangle$: Quantum state representing an individual.
    \item $|0\rangle$: Initial state of the fitness register.
    \item $|f(u)\rangle$: State encoding the fitness value of $u$, determined by the fitness function $f$.
\end{itemize}

The $U_{\text{fit}}$ operator is implemented using Controlled-NOT (CNOT) and Toffoli gates. It uses $n$-qubit register to represent the individual solution $|u\rangle$ and $m$-qubit fitness register initialized to $|0\rangle$. Auxiliary qubits are used to encode the fitness function in the $m$-qubit fitness register as a Boolean logic circuit.

\subsection{Oracle Subcircuit}
The oracle subcircuit marks high-fitness solutions by using a boolean function and phase-kickback to evaluates and flip the phase of states. Afterwards, those marked states will be amplified by Grover's Algorithm.\\

The Oracle circuit is implemented to perform the transformation:
\[
O: |u\rangle \otimes |f(u)\rangle \to (-1)^{g(f(u), T)} |u\rangle \otimes |f(u)\rangle,
\]
where:
\begin{itemize}
    \item \( |u\rangle \): Represents the quantum state encoding an individual.
    \item \( |f(u)\rangle \): Represents the quantum state encoding the fitness value of the individual.
    \item \( T \): A predefined threshold value, which for the MaxCut is the $|E|$
    \item \( g(f(u), T) \): A Boolean function that evaluates to 1 if \( f(u) > T \) and 0 otherwise.
\end{itemize}
This transformation marks the states with fitness values greater than the threshold \( T \) by applying a phase flip. The maximum possible fitness corresponds to the scenario where all edges are cut (a bipartite graph), making the fitness value equal to the total number of edges in the graph. By setting the threshold to the number of edges, the algorithm ensures that the search is confined to valid configurations, where each edge contributes to the fitness.

\subsubsection{The Quantum Adder}
To evaluate whether \( f(u) > T \), a quantum ripple-carry adder proposed by \cite{cuccaro2004newquantumripplecarryaddition} is utilized. The quantum adder operates reversibly and proceeds through a sequence of controlled operations as follows:
\[
|a\rangle \otimes |b\rangle \otimes |c_0\rangle \to |a\rangle \otimes |S\rangle \otimes |c_n\rangle,
\]
where \( |a\rangle \) represents the fitness value \( f(u) \), \( |b\rangle \) represents the threshold value \( T \), and \( |c_0\rangle \) is an ancillary qubit. The adder computes the sum \( S = f(u) + T \) modulo \( 2^n \), storing the result in \( |b\rangle \), while the carry bit \( c_n \) is stored in the ancillary qubit \( |c_n\rangle \). The operation is performed using a sequence of \textit{MAJ} (majority) gates to compute carry bits and \textit{UMA} (UnMajority and Add) gates to compute the sum and reverse intermediate changes. 

The adder evaluates whether \( f(u) > T \) by checking the carry qubit or the most significant bit (MSB) of the result, which indicates whether an overflow occurred during addition. This method ensures a reversible computation, adhering to the principles of quantum mechanics.

The Oracle circuit, incorporating this quantum ripple-carry adder guides the QGA optimization process. By comparing \( f(u) \) and \( T \) and marking individuals with high fitness, the Oracle ensures that genetic operations focus on promising solutions, accelerating convergence to the optimal result. The combination of the ripple-carry adder and reversible logic is particularly effective for solving the MaxCut, as it efficiently evaluates and identifies optimal solutions.

\subsection{Divide-and-Conquer Heuristic for QGA}
Here it is outlined the implementation of a divide-and-conquer heuristic for the Quantum Genetic Algorithm (QGA), inspired by its application in Quantum Approximate Optimization Algorithms (QAOA) \cite{zhou2023}. The heuristic uses graph partitioning, local optimization, and solution merging to solve large-scale problems, specifically the MaxCut problem, efficiently within the constraints of NISQ quantum hardware.\\

\textbf{Graph Partitioning:}
The goal is to divide the input graph \( G(V, E) \) into smaller subgraphs \( \{G_i(V_i, E_i)\} \), where each subgraph \( G_i \) has the number of vertices \( |V_i| \leq n \). The parameter \( n \) represents the maximum number of vertices that can fit in the quantum register used in the Quantum Genetic Algorithm (QGA).
\begin{itemize}
    \item \textit{Graph Representation:} The input graph \( G \) is represented by its set of vertices \( V \) and edges \( E \). For large graphs, directly encoding all vertices in the quantum register is infeasible for NISQ devices, necessitating a partitioning process.
    \item \textit{Initial Division:} Divide \( G \) into subgraphs \( \{G_i\} \) such that each \( G_i \) has fewer than or equal to \( n \) vertices. These subgraphs are created to balance size and structure, ensuring they are manageable within the constraints of the quantum hardware.
    \item \textit{Recursive Partitioning:} After the initial partitioning, some subgraphs may still exceed the quantum register's size limit \( n \). Recursive partitioning ensures that all subgraphs eventually satisfy the size constraint.
    \item \textit{Condition Check:} For each subgraph \( G_i \), check if \( |V_i| > n \). If so, apply the partitioning process again to \( G_i \).
    \item \textit{Divide Further:} Break the oversized subgraph into smaller subgraphs using iterative techniques to ensure size constraints are met.
    \item \textit{Repeat Until Completion:} The process continues recursively until all subgraphs \( G_i \) satisfy \( |V_i| \leq n \).
\end{itemize}
Although this heuristic can be helpful in terms of scalability and it allow us to perform tests on NISQ-era machines, it also has its limitations since it lacks accuracy by losing boundary edges between subgraphs, and the necessity to be recursively implemented for large graphs.\\
\textbf{Practical Limitations}
\begin{itemize}
    \item \textbf{Boundary Edges:} During partitioning, edges connecting vertices in different subgraphs are known as boundary edges. These edges require careful handling, especially when applying algorithms like MaxCut, as their inclusion/exclusion affects the overall optimization.
    \item \textbf{Subgraph Characteristics:} The partitioning process often seeks to create subgraphs that are not only small enough to fit within the quantum register but also retain meaningful structures.
    \item \textbf{Recursive Depth:} The depth of recursion is determined by the initial graph size and the register limit \( n \). For very large graphs, this could result in multiple levels of nested subgraphs.
\end{itemize}

The graph \( G \) is represented as a set of disjoint or overlapping subgraphs \( \{G_i(V_i, E_i)\} \), each of which satisfies the condition \( |V_i| \leq n \).
 These smaller subgraphs can then be processed independently within the constraints of the quantum register, enabling for the optimization of the MaxCut problem.

\subsection*{Local Optimization with the QGA}
\begin{enumerate}
    \item \textbf{Encode Subgraphs:}
    \begin{itemize}
        \item Represent individuals as quantum states in a superposition.
        \item Use fitness evaluation circuits tailored to the MaxCut problem to calculate cuts for each subgraph.
    \end{itemize}

    \item \textbf{Apply QGA:}
    \begin{itemize}
        \item Perform selection, crossover, and mutation operations on the quantum register.
        \item Compute fitness values without intermediate measurements by leveraging superposition.
        \item Terminate when the stopping Complexitycriterion (\textit{e.g}., iteration count or convergence) is met.
    \end{itemize}

    \item \textbf{Store Subgraph Solutions:}
    Save the optimized solutions \( \{x_i\} \) for each subgraph.
\end{enumerate}

\subsection*{Solution Merging}
\begin{enumerate}
    \item \textbf{Reformulate the Problem:}
    \begin{itemize}
        \item Treat each subgraph as a vertex in a new meta-graph \( G'(V', E') \).
        \item Assign weights to edges in \( G' \) based on connectivity and edge weights between subgraphs in the original graph \( G \).
        \item Handle symmetry (\textit{e.g.}, \( \mathbb{Z}_2 \)) by considering both \( x_i \) and its complement \( \bar{x_i} \) as valid solutions for each subgraph.
    \end{itemize}

    \item \textbf{Solve the Meta-Graph:}
    \begin{itemize}
        \item Use QGA to find the optimal cut for \( G' \).
        \item Interpret the meta-graph solution to determine the global cut for \( G \).
    \end{itemize}
\end{enumerate}

\subsection*{Recursive Refinement}
If the meta-graph \( G' \) exceeds the size limit \( n \):
\begin{enumerate}
    \item \textbf{Recursive Application:}
    \begin{itemize}
        \item Partition \( G' \) into smaller subgraphs.
        \item Optimize the subgraph solutions using QGA.
        \item Repeat until the size of the final meta-graph is manageable within the quantum register's capacity.
    \end{itemize}
\end{enumerate}

\subsection{Complexity Analysis}
\label{sec:complexity}

The complexity analysis of the Algorithm for the MaxCut problem accounts for the inherent challenges of simulating quantum algorithms on classical computers.) the simulation of quantum circuits requires exponential runtime with respect to the circuit size. 
The total number of qubits required for QGA in the MaxCut problem is determined by the function \cite{udrescu2006}:

\[
f(|V|, n, M, m) = |V| \cdot n + 2 \cdot (M + m) + 3,
\]

where:
\begin{itemize}
    \item \( |V| \): The number of vertices in the graph.
    \item \( n \): The number of qubits required to encode vertex states or partitions.
    \item \( M \): The number of qubits needed to represent the fitness value in two's complement, typically \( \lceil \log_2(E) \rceil \), where \( E \) is the number of edges in the graph.
    \item \( m \): The number of Grover iterations, calculated as \( m = O(\sqrt{2^M}) \).
    \item Additional qubits: 3 qubits are used for carry-in, oracle workspace, and a validity flag, while Grover's adders require 2 qubits for carry-out during each iteration.
\end{itemize}

Remembering that a complete graph is a upper-bound of the number of edges for any graph witn $n$ vertices, we can derivate a function,
\[
g(n) = n^2 + 2^{\frac{n(n-1)}{4}+1} + 3,
\]
\
Which bounds the $f$ and give us the minimum amount of qubits necessary to run a graph with $n$ vertices. Due to the exponential growth of required resources with \( |V| \), \( M \), and \( m \), the scalability of the algorithm in simulation is limited. These constraints underscore the necessity of access to actual quantum hardware to fully exploit the potential of the application for large graph instance without the downside of a divide-and-conquer heuristic. Despite these simulation challenges, the theoretical framework of QGA retains its quantum advantage, offering an efficient approach to solve combinatorial optimization problems like MaxCut.

\section{Experiments}
\label{sec:experiments}

The QGA for the MaxCut has been evaluated experimentally on various classes of graphs, including complete graphs and Erd\H{o}s-R\'{e}nyi random graphs, to demonstrate its effectiveness in solving the MaxCut problem. It is presented a comparison of the QGA with the Semidefinite Programming (SDP) approach and highlights its performance on specific graph types. All experiments were conducted under similar conditions to ensure the validity of comparisons.\\
The measurements were conducted by implementing the algorithm using Qiskit, with simulations performed on the IBM quantum platform. The $ibmq\_qasm\_simulator$ backend, provided by the $ibm-q$ provider, was utilized. This simulator is a versatile, context-aware tool capable of simulating quantum circuits under ideal conditions or with noise modeling, supporting circuits with up to 29 qubits. The code can be found here: \href{https://github.com/pauloaviana/maxcut-qga}{https://github.com/pauloaviana/maxcut-qga}

\subsection{Setup}
The experiments were conducted on two types of graphs on a Qiskit Simulator:
\begin{itemize}
    \item \textbf{Complete Graphs:} Graphs where every pair of vertices is connected by an edge.
    \item \textbf{Erd\H{o}s-R\'{e}nyi Random Graphs:} Graphs generated with a fixed probability for edge inclusion, denoted as $G(n, p)$.
\end{itemize}

For each graph type, the performance of QGA and SDP was compared based on the cut values achieved. Results were averaged over multiple runs to account for randomness inherent in both algorithms.

\subsection{Results for Small Graphs}
For graphs small enough to run the QGA directly, those without a divide-and-conquer heuristic (which by itself loses boundary edges), are limited up to $|V| = 8$ vertices. Both the the SDP and the QGA got the optimal result for graphs (complete or Randomly generated) up to such size.

\subsection{Results for Complete Graphs}
The experiments on complete graphs showed that the QGA consistently found the true optimal MaxCut values, which are determined by $\left\lfloor \frac{n^2}{4} \right\rfloor$, while the SDP approach achieved approximate results withing the theoretical limit $0.878$, Table \ref{tab:complete_graphs} summarizes these results.

\begin{table}[h!]
    \centering
    \caption{Comparison of QGA and SDP on Complete Graphs.}
    \label{tab:complete_graphs}
    \begin{tabular}{@{}ccccc@{}}
        \toprule
        \textbf{Number of Vertices} & \textbf{QGA (Optimal Value)} & \textbf{SDP Value} & \textbf{QGA Ratio} & \textbf{SDP Ratio} \\
        \midrule
        3 & 2 & 2 & 1.0 & 1.0 \\
        5 & 6 & 6 & 1.0 & 1.0 \\
        8 & 16 & 15 & 1.0 & 0.9375 \\
        12 & 36 & 35 & 1.0 & 0.9722 \\
        23 & 132 & 130 & 1.0 & 0.9848 \\
        31 & 240 & 237 & 1.0 & 0.9875 \\
        56 & 784 & 780 & 1.0 & 0.9949 \\
        80 & 1600 & 1593 & 1.0 & 0.9969 \\
        128 & 4096 & 4085 & 1.0 & 0.9973 \\
        \bottomrule
    \end{tabular}
\\The table shows that QGA proposted got the true value for the MaxCut for all the tested instances, while the SDP slowly gets worst results as the number of vertices grows, as expected.
\end{table}

\subsection{Results for Erd\H{o}s-R\'{e}nyi Random Graphs}
For Erd\H{o}s-R\'{e}nyi random graphs, the performance of QGA varied depending on the specific instance, but it consistently demonstrated competitive or superior results compared to SDP. Two separate tables, Tables \ref{tab:erdos_renyi_qga1} and \ref{tab:erdos_renyi_qga2} respectivelly show the results for the median value found during the runs and for the best result found for the QGA. A comparison of the performance of QGA runs and SDP values is made, including the ratio of QGA results to SDP values.

\begin{table}[h!]
    \centering
    \caption{Comparison of QGA Run 1 and SDP on Erd\H{o}s-R\'{e}nyi Random Graphs.}
    \label{tab:erdos_renyi_qga1}
    \begin{tabular}{@{}cccc@{}}
        \toprule
        \textbf{Graph Instance} & \textbf{QGA Med.} & \textbf{SDP Value} & \textbf{QGA/SDP Ratio} \\
        \midrule
        $G(50, 0.1)$ & 91 & 92 & 0.9891 \\
        $G(50, 0.25)$ & 194 & 210 & 0.9238 \\
        $G(50, 0.5)$ & 346 & 360 & 0.9611 \\
        $G(50, 0.75)$ & 478 & 524 & 0.9122 \\
        $G(100, 0.1)$ & 232 & 329 & 0.7052 \\
        $G(100, 0.25)$ & 674 & 786 & 0.8576 \\
        $G(100, 0.5)$ & 1297 & 1361 & 0.9529 \\
        $G(100, 0.75)$ & 1894 & 2016 & 0.9394 \\
        $G(200, 0.1)$ & 1017 & 1211 & 0.8401 \\
        $G(200, 0.25)$ & 2550 & 2778 & 0.9180 \\
        $G(200, 0.5)$ & 5095 & 5326 & 0.9566 \\
        $G(200, 0.75)$ & 7494 & 7815 & 0.9589 \\
        $G(350, 0.1)$ & 3120 & 3611 & 0.8640 \\
        $G(350, 0.25)$ & 7771 & 8236 & 0.9436 \\
        $G(350, 0.5)$ & 15443 & 16030 & 0.9634 \\
        $G(350, 0.75)$ & 22941 & 23530 & 0.9749 \\
        $G(500, 0.1)$ & 6335 & 7097 & 0.8926 \\
        $G(500, 0.25)$ & 15684 & 16520 & 0.9493 \\
        $G(500, 0.5)$ & 31316 & 33110 & 0.9456 \\
        $G(500, 0.75)$ & 46875 & 48130 & 0.9740 \\
        \bottomrule
        \bottomrule
    \end{tabular}
\\This table takes one average run of the QGA and compares the result with the SDP, The results are competitive to the SDP and since there are lost boundary edges using the graph contraction approach, the results obtained are inferior to an expected application with more qubits.
\end{table}

\begin{table}[h!]
    \centering
    \caption{Comparison of QGA Best runs and SDP on Erd\H{o}s-R\'{e}nyi Random Graphs.}
    \label{tab:erdos_renyi_qga2}
    \begin{tabular}{@{}cccc@{}}
        \toprule
        \textbf{Graph Instance} & \textbf{QGA Best} & \textbf{SDP Value} & \textbf{QGA/SDP Ratio} \\
        \midrule
        $G(50, 0.1)$ & 96 & 92 & 1.0435 \\
        $G(50, 0.25)$ & 240 & 210 & 1.1429 \\
        $G(50, 0.5)$ & 320 & 360 & 0.8889 \\
        $G(50, 0.75)$ & 512 & 524 & 0.9771 \\
        $G(100, 0.1)$ & 343 & 329 & 1.0426 \\
        $G(100, 0.25)$ & 783 & 786 & 0.9962 \\
        $G(100, 0.5)$ & 1375 & 1361 & 1.0103 \\
        $G(100, 0.75)$ & 2024 & 2016 & 1.0040 \\
        $G(200, 0.1)$ & 1250 & 1211 & 1.0322 \\
        $G(200, 0.25)$ & 2861 & 2778 & 1.0299 \\
        $G(200, 0.5)$ & 5423 & 5326 & 1.0182 \\
        $G(200, 0.75)$ & 7875 & 7815 & 1.0077 \\
        $G(350, 0.1)$ & 3639 & 3611 & 1.0078 \\
        $G(350, 0.25)$ & 8583 & 8236 & 1.0421 \\
        $G(350, 0.5)$ & 16030 & 16030 & 1.0000 \\
        $G(350, 0.75)$ & 23740 & 23530 & 1.0089 \\
        $G(500, 0.1)$ & 7034 & 7097 & 0.9911 \\
        $G(500, 0.25)$ & 17140 & 16520 & 1.0375 \\
        $G(500, 0.5)$ & 33140 & 33110 & 1.0009 \\
        $G(500, 0.75)$ & 48200 & 48130 & 1.0015 \\
        \bottomrule
    \end{tabular}
\\The results shown are the best picks after multiple runs, it's possible to see slightly better results even with the disadvantage of the graph contraction heuristic. Such results are not conclusive but they show a potential advantage of the QGA over the SDP even with a low number of qubits available.
\end{table}

\section{Analysis and Discussion}
\label{sec:analysis}

The QGA consistently achieved the true MaxCut values for all tested complete graphs. This comes from the fact that for complete graphs you do not lose edges when partitioning it with the implemented heuristic due to its binary symmetry. This shows that the QGA can consistently optimize for the subgraphs and their weighted version when recursively partitioned. In contrast, the SDP approach occasionally underperformed as the number of vertices go up, as it's expected from it. The same situation happened for the small graphs tested, those that have up to 8 vertices, where in this case both the QGA and the SDP found the optimal solutions

For Erd\H{o}s-R\'{e}nyi graphs, the QGA displayed variability in results across multiple runs, likely due to the nature of  probabilistic measurements. However, it frequently outperformed SDP when selected for the best result run found. Moreover, even when taken on average, the QGA yields a good performance in comparison with the SDP, producing results up to 96\% of the solution found by the SDP itself.

\section{Conclusion}
\label{sec:conclusion}

The experimental results demonstrate the effectiveness of QGA with a divide-and-conquer heuristic in solving the MaxCut problem, especially for small and complete graphs, where it consistently found the true best cut value. For Erd\H{o}s-R'{e}nyi random graphs, even with the disadvantage of using the divide-and-conquer heuristic, which intrinsically loses boundary edges, QGA often achieved competitive or superior results, making it a promising quantum framework for complex optimization problems in the NISQ era and beyond.

For future work, further testing would help demonstrate the applicability and limitations of the proposed algorithm, particularly by utilizing a NISQ machine with a sufficient number of logical qubits to execute it. Refining quantum circuit components, such as the fitness operator, oracle, and Grover diffusion operator, could minimize qubit requirements and enhance performance. The usage of hybrid quantum-classical optimization techniques, such as Variational Quantum Circuits (VQCs), could be explored. These methods utilize parameterized quantum circuits and iterative updates to find optimal solutions, offering an alternative to Grover's algorithm. Specifically, the Grover oracle and diffusion operator could be replaced with a variational cost function that directly encodes the problem constraints. The performance of a VQC optimizer could then be compared in terms of solution quality, convergence speed, scalability, circuit depth, number of required qubits. Adapting the QGA framework to address other combinatorial optimization problems would be a valuable direction for exploring its efficiency, or even for tackling broader problems traditionally handled by classical genetic algorithms. As quantum hardware continues to advance, incorporating weighted edges into the algorithm will become feasible, making it possible to adapt to problems like the Travelling Salesman Problem (TSP), which often includes weighted edges and a global constraint, this would test the framework's flexibility, especially when scaling up to larger problem instances. Additionally, the divide-and-conquer heuristic could be significantly improved with an increasing number of available qubits, allowing for larger partitioned subgraphs and enabling the processing of more complex graphs without losing crucial boundary edges. Applying the algorithm to real-world scenarios, such as circuit design or statistical physics, could demonstrate its practical utility while showing new challenges and opportunities for optimization. Finally, the framework could be adapted to other NP-hard graph problems to demonstrate its generalizability and practical efficiency.

\nocite{*}

\printbibliography
\end{document}